\documentstyle[11pt,newpasp,twoside,epsf]{article}
\begin{document}

\title{T Tauri Stars in the Small Magellanic Cloud}

\author{Valentin D. Ivanov}
\affil{European Southern Observatory, Ave. Alonso de Cordova 3107,
Santiago 19001, Chile; vivanov@eso.org}

\author{Ray Jayawardhana}
\affil{Department of Astronomy, University of Michigan, 830 Dennison
Building, Ann Arbor, MI 48109, U.S.A.; rayjay@astro.lsa.umich.edu}

\begin{abstract}
The Small Magellanic Cloud (SMC) is an excellent laboratory to study
the formation of solar-mass stars in a low-metallicity environment, similar
to the conditions expected in the early phases of galactic evolution.
Here we present preliminary results from a search for low-mass
pre-main-sequence stars in the SMC based on {\it Hubble Space Telescope}
archival data. Candidates are selected on the basis of their H$\alpha$
emission and location in the [(F675W-F814W), F814W] color-magnitude
diagram. We discuss characteristics of our candidate T Tauri sample and 
possible follow up work. 
\end{abstract}

\section{Introduction}

Young stellar populations often provide the most luminous and
prominent stars in galaxies. They trace the ongoing star
formation (SF hereafter), and are closely related to its
triggering mechanisms. Studies of young stars can help 
probe chemical enrichment processes of the Universe, and 
even find local analogs of high redshift systems.

Galactic SF studies take advantage of observations of
individual pre-main-sequence (PMS) stars in nearby young
clusters and associations. Unfortunately, only the most massive 
young stars can be observed in external galaxies; while these 
high-mass stars dominate the light from their host systems, they
account for only a fraction of the mass and part of the chemical
enrichment. The Magellanic Clouds are the sole exception, offering
the possibility to study the formation and early evolution of 
the most-common low-mass stars at metallicities of only $\sim1/5$ 
and $\sim1/20$ of the solar abundance. The Small Magellanic Cloud 
(SMC) provides the closest local analog to the conditions in which 
bulk of the SF at high redshift took place.

Candidate low-mass PMS stars have been detected in the Large 
Magellanic Cloud (LMC) by Panagia et al. (2000), based on their 
H$\alpha$ excess, in the region around Supernova 1987A. They 
used multi-color photometry to constrain the effective temperatures, 
to measure the IMF slope, and to study the spatial distribution of 
high- and low-mass PMS stars. Wichmann et al. (2001) reported the 
first spectroscopically confirmed extragalactic T Tauri star in the
LMC. Their sample for multi-object spectroscopy was chosen based on 
the near-infrared excess, and yielded one detection of H$\alpha$ 
emission from 19 candidates. The low success rate is somewhat 
surprising since circumstellar disks are expected to show both 
infrared excess and accretion signatures in the form of H$\alpha$ 
emission. Brandner et al. (2001) relied on the near-infrared excess 
alone to separate candidate young stars from the underlying populations 
in the 30 Doradus region of the LMC.

We have commenced a program to select T Tauri candidates in the SMC, 
in preparation for future spectroscopy. Here we discuss the selection 
technique, the sample for follow-up optical spectroscopy, and some 
preliminary results.

\section{Field Selection and Data Reduction}

The SMC was chosen as a primary target for this work because of
its lower metallicity in comparison with the LMC. The relative
proximity of the Magellanic Clouds allows us to resolve the
stellar populations easily with the $WFPC2$ on the $Hubble~ Space~
Telescope~ (HST)$. All stars are at roughly same distance, and the
Galactic extinction is low $\rm(A_B=0.17$ mag; Burstein \& Heiles 
1982).

We searched the $HST/WFPC2$ archive for H$\alpha$ images in the
vicinity of clusters in the catalog of Bica \& Dutra (2000),
ensuring the presence of young stellar population in the fields.
We selected the fields with total integration in H$\alpha$
longer than 1000 sec, and with sufficiently long F675W and F814W
exposures. The F675W observations are needed for proper continuum
subtraction, and a second broad-band filter is used to construct
a color-magnitude diagram (CMD).

A single SMC field near the open cluster Kron\,57
$(\alpha$=1:08:12, $\delta$=$-$73:14:38, Eq. J2000) satisfies
all our criteria. The data were obtained as part of a 
program to study the extended gas emission (P.I. Garnet, ID 8196). The
total integration times (always split in three separate exposures)
are 3000, 600, and 1200, in F656N, F675W, and F814W, respectively.

Aperture photometry was performed on the median-combined
images, and was calibrated following the prescription of
Holtzman et al. (1995a, 1995b). This procedure is adequate since 
the field does not suffer from severe crowding.

\section {Selection Criteria for PMS Candidates}

The PMS candidates were selected on the basis of two criteria.
The primary one was a reliable detection of H$\alpha$ excess
emission. We measured all sources in the F656N and F675W images,
and obtained their magnitudes. Then, we imposed a condition for
``strong'' excess, following the example of Panagia et al. (2000):
$m({\rm F675W})-m({\rm F656N})\geq0.3$ mag and both measurements
are at least $4\sigma$ detections, where $\sigma$ is the
$IRAF/APPHOT$ error, obtained with proper gain factors and readout
noise for the corresponding $WFPC2$ chips. The limit above
corresponds to H$\alpha$ equivalent width of about 8\,\AA~ or
larger (see Biretta 1996 for the $HST/WFPC2$ filter parameters).
The spatial locations of the ``strong'' H$\alpha$ emitters are
shown in Figure 1 (left).

The secondary condition was the position of the candidates on the
[(F675W-F814W),\,F814W] CMD. It eliminates from the sample bright
Be stars with $m({\rm F656N})\sim15-18$ mag (see also the
discussion in Sec. 5 of Panagia et al. 2000). A few such Be stars 
were detected by Panagia et al (2000), and we have one in our field, 
on $WF2$ (see Figure 1, right). The various sequences on the diagrams 
are somewhat blurred by the reddening variation and the mixture of
stellar populations with different ages and metallicities. The CMD
also allows us to exclude some RGB/AGB stars with emission due to
extended envelopes and mass loss.

\begin{figure}
\plottwo{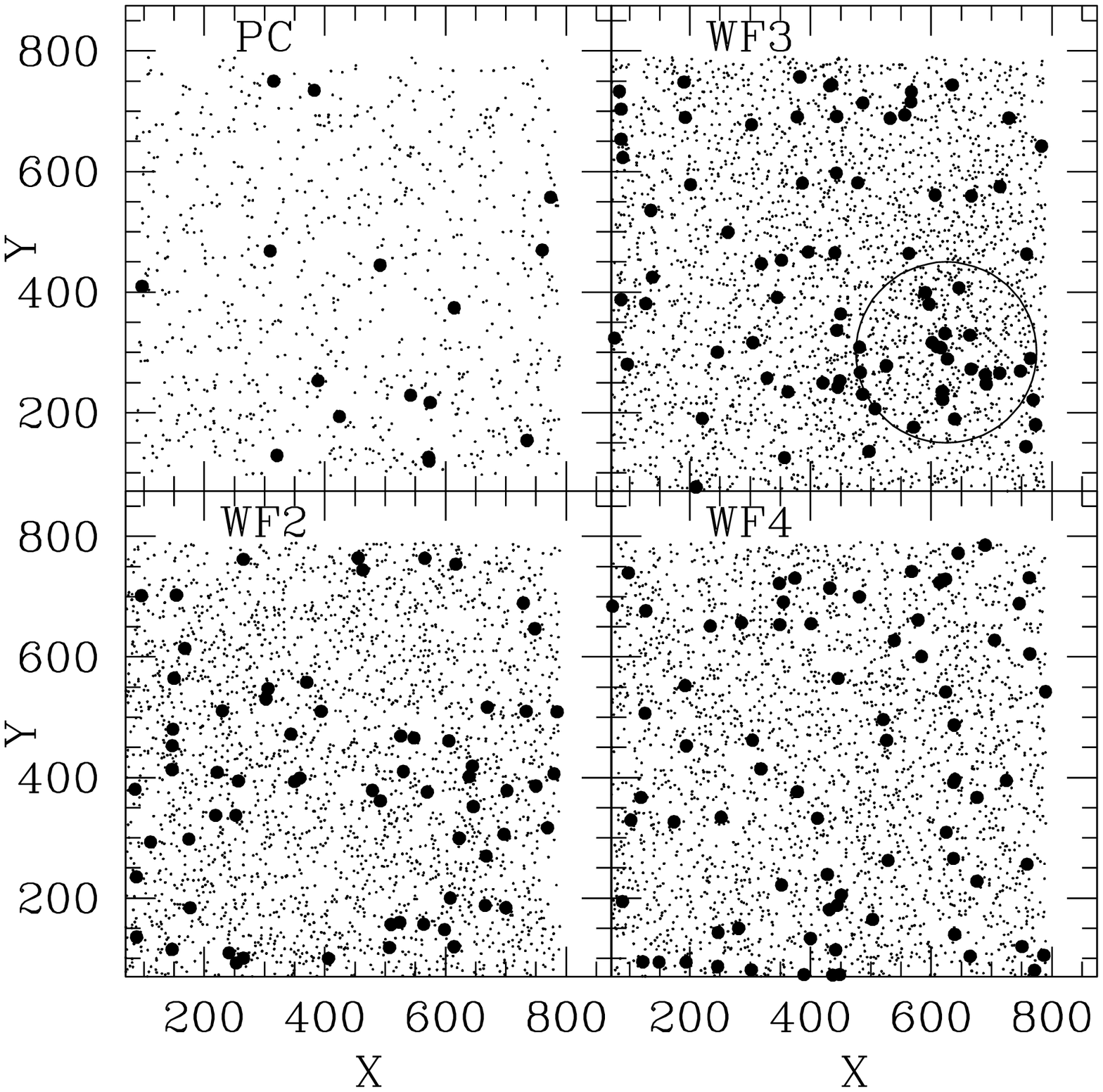}{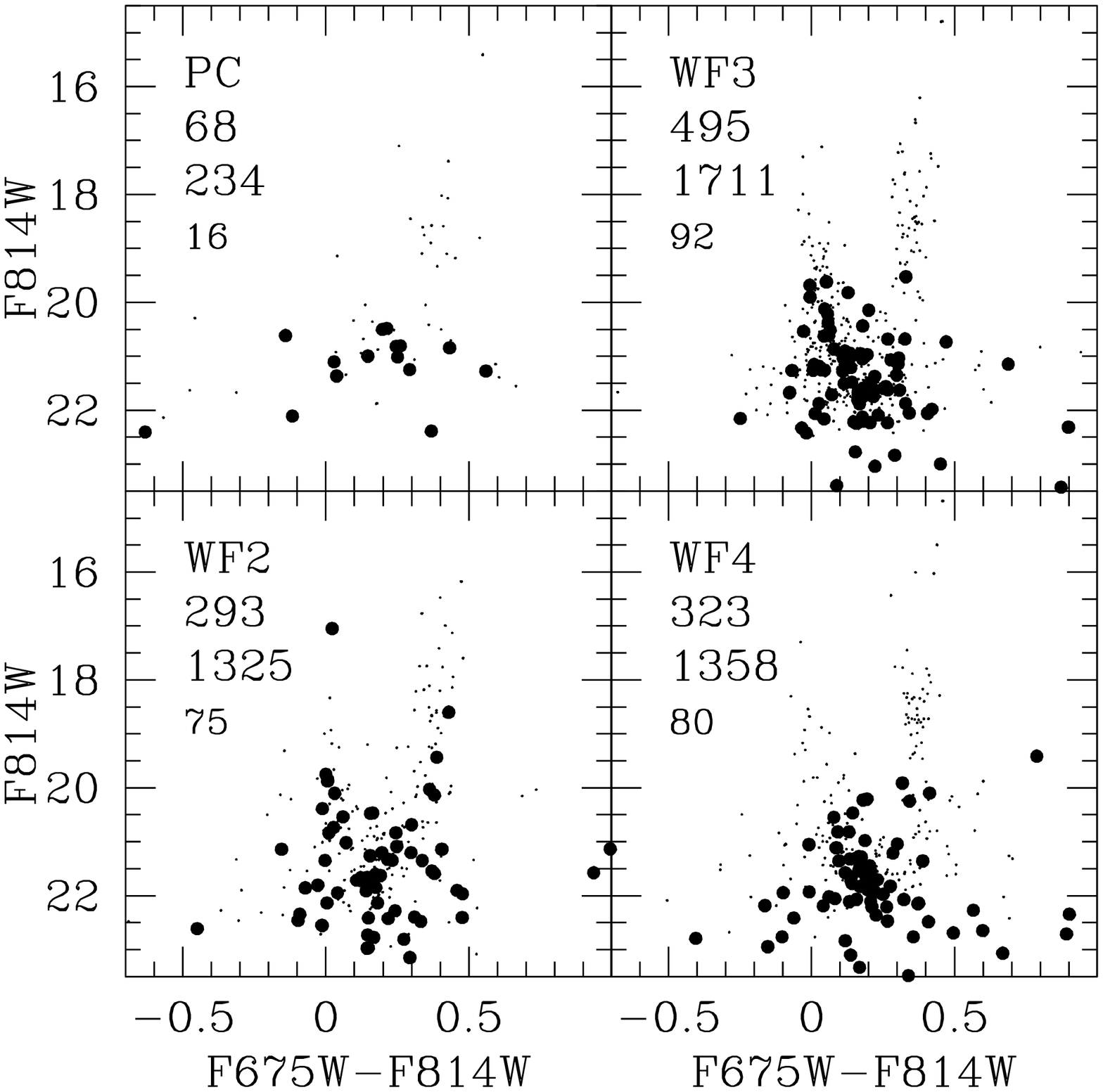}
\caption{Stellar sample. All stars detected in F814W are
indicated with small dots, and stars with $4\sigma$ H$\alpha$
excess are solid dots.
$Left:$ Spatial distribution. The big circle on WF3
represents the position of Kron\,57.
$Right:$ Color-magnitude diagram. For each chip are given:
number of stars detected in H$\alpha$, number of stars
with F675W-F814W, and number of stars with $4\sigma$
H$\alpha$ excess.}
\end{figure}

\section{Discussion}

We have measured a total of 4628 stars in F675W and F814W. Only 1179
of them were detected on the F656N image, and 263 of these passed
the ``strong'' H$\alpha$ emitter criterion discussed in the
previous section. About 10\% were classified as Be or RGB/AGB stars,
or were excluded as outliers on the CMD, leaving us with $\sim$240
PMS candidates. The ``color-color'' diagrams
[(F675W-F814W),(F675W-F656N)] are shown in Figure 2 (left). The
surface density of PMS candidates follows the overall
distribution of stars in the field, with an increase toward the
center of Kron\,57.

The histograms of H$\alpha$ equivalent width are shown in Figure 2
(right). They resemble closely those for the SN\,1987A region
(Panagia et al. 2000), and feature no significant differences
between Kron\,57 and the field, indicating that the latest
generation of stars in the cluster and in the field may have formed
approximately at the same time, and with a similar IMF. Indeed, the
fraction of ``strong'' H$\alpha$ emitters is constant within
the uncertainties in all chips (5.4-5.9\%$\pm$0.5\% for the WFs;
6.8\%$\pm$1.7\% for the PC). However, this conclusion requires
further verification with a larger sample and more colors. The
H$\alpha$ emission properties of our SMC PMS candidates differ
significantly from the Galactic sample (see Figure 5 in Panagia
et al. 2000), as do the T Tauri candidates in the vicinity of
SN\,1987A in the LMC. Most probably, this is due to an age
difference between young stars in the Magellanic Clouds and in 
the Milky Way.

The adopted selection criteria introduce a bias toward 
(actively accreting) ``strong-line'' T Tauri stars. This is an
unfortunate consequence of the desire to improve the success rate
for follow-up spectroscopy. An alternative alleviating this
bias is to use the variability of T Tauri stars (Brice\~{n}o et al.
2001) for candidate selection. It is a time consuming method but 
the microlensing surveys toward the Magellanic Clouds provide a 
useful data set (Lamers et al. 1999). Our final PMS sample can still 
be compared with the subset of Galactic ``strong-line'' T Tauri 
stars (e.g. from Brice\~{n}o et al. 2001) if appropriate
constraints are applied to the latter.

\begin{figure}
\plottwo{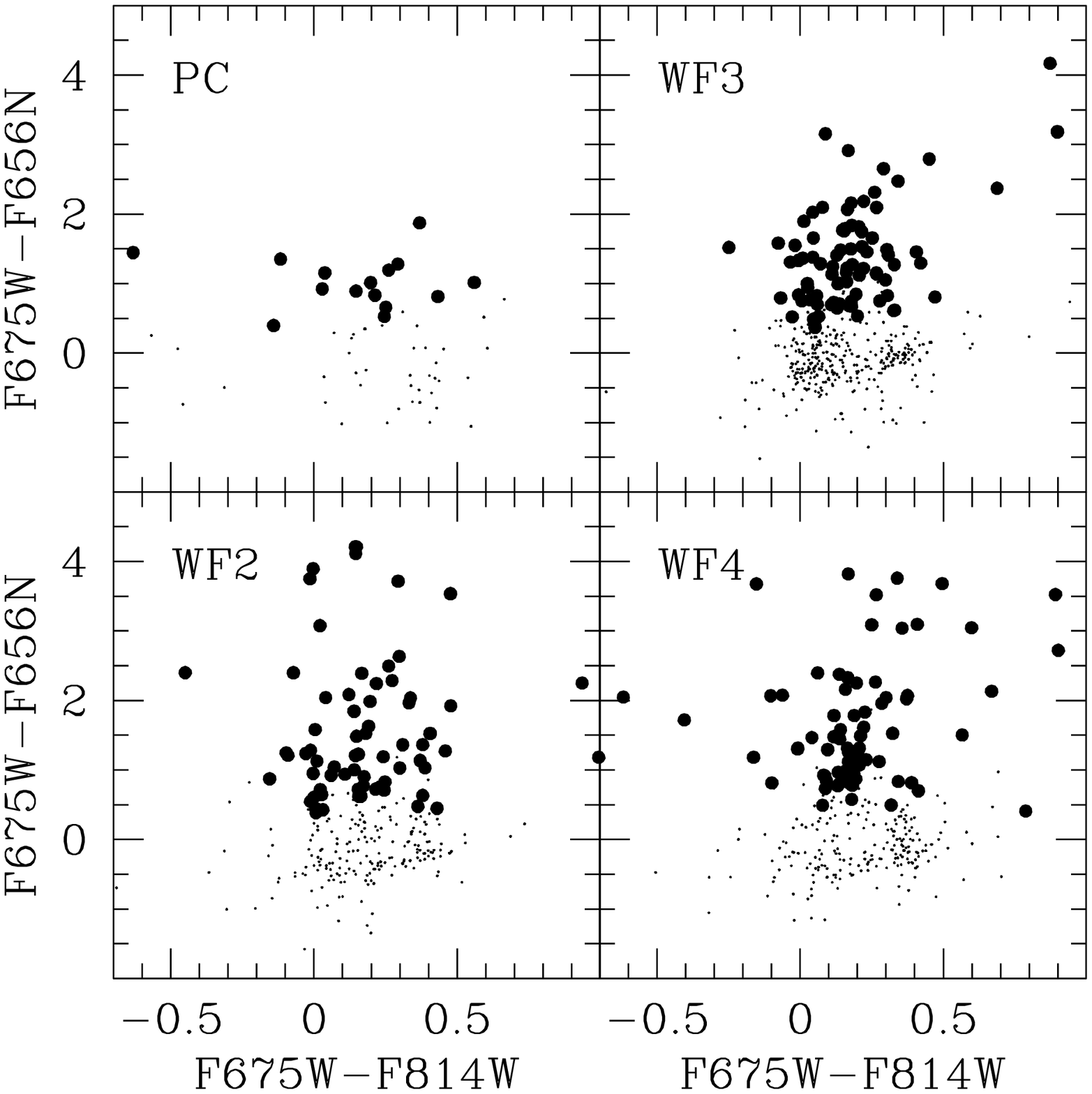}{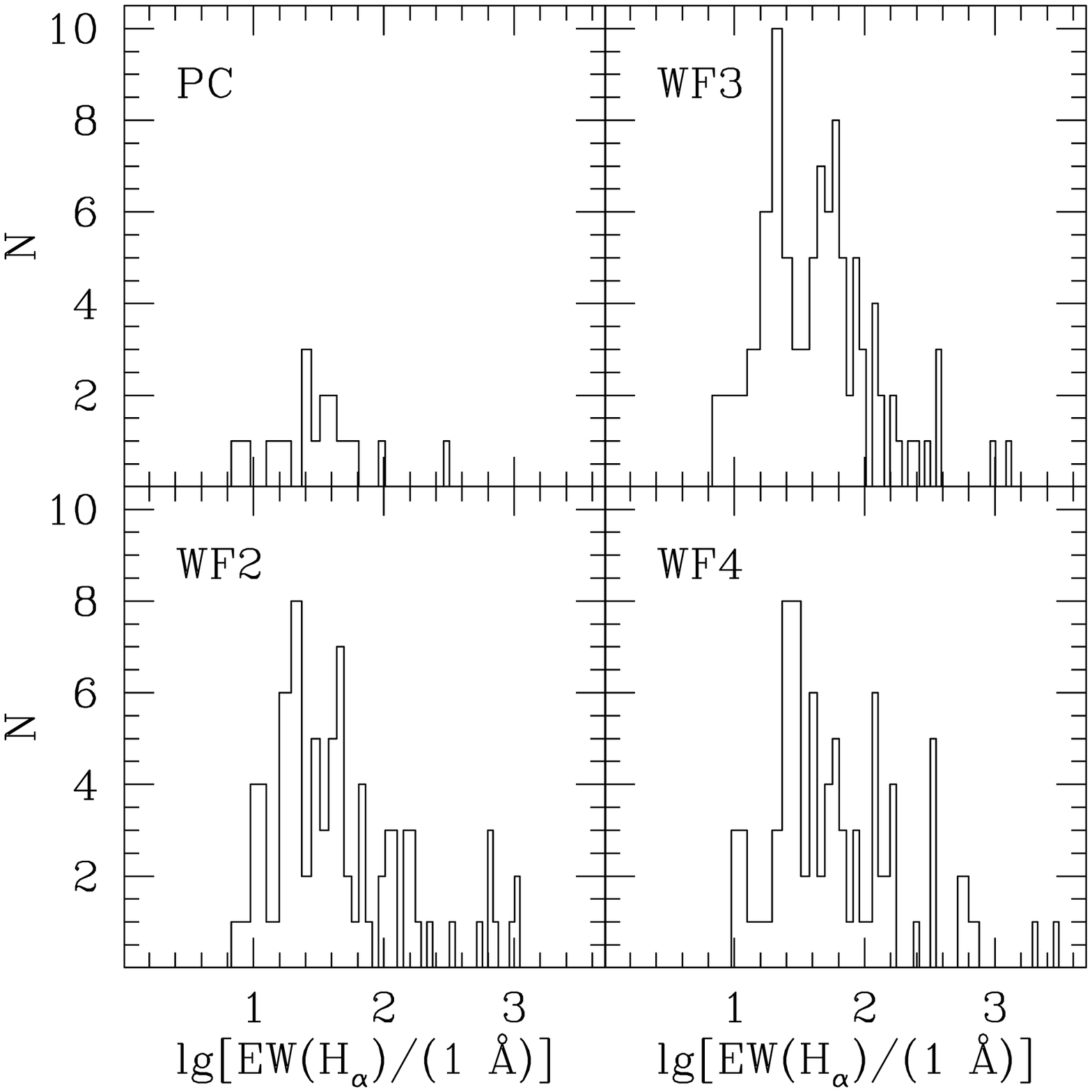}
\caption{Stellar sample. See Figure 1 for the details.
$Left:$ Color-color diagram.
$Right:$ Histogram of the H$\alpha$ equivalent width.}
\end{figure}

\section{Results}

\begin{enumerate}
\item Candidate PMS stars in the SMC are detected for
the first time, based on their H$\alpha$ excess.

\item The locus of selected candidates on the CMD is consistent
with them being PMS stars.

\item The distribution of equivalent widths is consistent with
that of Panagia et al. (2000) for the PMS candidates in
the region around SN\,1987A in the LMC. This distribution is
different from that for Galactic strong-line T Tauri stars,
suggesting a possible age difference.
\end{enumerate}

\newpage
\acknowledgments
RJ is grateful to ESO for a Visiting Scientist position in Santiago,
Chile which made this collaboration possible. All of the data presented 
in this paper were obtained from the
Multimission Archive at the Space Telescope Science Institute
(MAST). STScI is operated by the Association of Universities
for Research in Astronomy, Inc., under NASA contract NAS5-26555.
Support for MAST for non-HST data is provided by the NASA Office
of Space Science via grant NAG5-7584 and by other grants and
contracts.


\begin{references}
\reference Bica, E. \& Dutra, C.M. 2000, \aj, 119, 1214
\reference Biretta, J.A. 1996, $WFPC2$ Instrument Handbook 4.0
	(Baltimore: STScI)
\reference Brice\~{n}o, C., Vivas, K.A., Calvet, N., Hartmann, L.,
	Pacheco, R., Herrera, D., Romero, L., Berlind, P.,
	S\'{a}nchez, G., Snyder, J.A., \& Andrews, P. 2001, Science,
	291, 93
\reference Brandner, W., Grebel, E.K., Barb\'{a}, R.H., Walborn,
	N.R., \& Moneti, A. 2001, \aj, 122, 858
\reference Burstein, D. \& Heiles, C. 1982, \aj, 87, 1165
\reference Holtzman, J.A., Burrows, C.J., Casertano, S., Hester,
	J.J., Trauger, J.T., Watson, A.M., \& Worthey, G. 1995a,
	\pasp, 107, 1065
\reference Holtzman, J.A., Hester, J.J., Casertano, S., et al.
	1995b, \pasp, 107, 156
\reference Lamers, H.J.G.L.M., Beaulieu, J.P., \& de Wit, W.J.
	1999, \aap, 341, 827
\reference Panagia, N., Romanielo, M., Scuderi, S., \& Kirshner,
	R.P. 2000, \apj, 539, 197
\reference Wichmann, R., Schmitt, J.H.M.M., \& Krautter, J. 2001,
	\aap, 380, L9
\end{references}
\end{document}